\documentclass[12pt]{article}
\usepackage{latexsym}
\usepackage{amsthm}
\usepackage{amsmath,amssymb,amsfonts,bm,amsbsy,bbm,mathtools,mathabx}
\usepackage{graphicx,rotating,lscape}
\usepackage{verbatim}
\usepackage{natbib}
\usepackage{multirow,color}
\usepackage{geometry}
\usepackage{setspace}
\usepackage{subfigure}
\usepackage{float}
\usepackage{booktabs}

\usepackage{enumitem}
\usepackage{authblk}

\newtheorem{lemma}{Lemma}
\newtheorem{theorem}{Theorem}
\newtheorem{corollary}{Corollary}
\newtheorem{proposition}{Proposition}

\newtheorem{remark}{Remark}
\newtheorem{assumption}{Assumption}

\newcommand{\indep}{\rotatebox[origin=c]{90}{$\models$}}

\def\A{{\bf A}}

\def\B{{\bf B}}

\def\f{{\bf f}}

\def\h{{\bf h}}

\def\X{{\bf X}}
\def\x{{\bf x}}

\def\u{{\bf u}}

\def\calT{{\cal T}}

\def\bphi{\boldsymbol\phi}
\def\bvarphi{\boldsymbol\varphi}

\def\bmu{\boldsymbol\mu}
\def\btau{\boldsymbol\tau}

\def\bmu{\boldsymbol\mu}

\def\bdelta{{\boldsymbol\delta}}

\def\0{{\bf 0}}
\def\trans{^{\rm T}}
\def\pr{\hbox{pr}}
\def\wh{\widehat}
\def\wt{\widetilde}

\def\var{\hbox{var}}

\def\eff{_{\rm eff}}

\def\n{\nonumber}

\def\log{{\rm log}}

\def\squarebox#1{\hbox to #1{\hfill\vbox to #1{\vfill}}}

\def\bse{\begin{eqnarray*}}
	\def\ese{\end{eqnarray*}}
\def\be{\begin{eqnarray}}
	\def\ee{\end{eqnarray}}
\def\bsq{\begin{equation*}}
	\def\esq{\end{equation*}}
\def\bq{\begin{equation}}
	\def\eq{\end{equation}}

\def\fx{f_{\X}}

\def\sumi{\sum_{i=1}^n}

\def\sumIP1{\sum_{i=1, i\in P_1}^N}

\def\boxit#1{\vbox{\hrule\hbox{\vrule\kern6pt\vbox{\kern6pt#1\kern6pt}\kern6pt\vrule}\hrule}}

\definecolor{mygreen}{RGB}{34, 139, 34} 

\usepackage{tikz}
\usetikzlibrary{positioning,shapes.geometric}

\makeatletter
\newcommand*{\ind}{%
	\mathbin{%
		\mathpalette{\@ind}{}%
	}%
}
\newcommand*{\nind}{%
	\mathbin{
		\mathpalette{\@ind}{\not}
	}%
}
\newcommand*{\@ind}[2]{%
	\sbox0{$#1\perp\m@th$}
	\sbox2{$#1=$}
	\sbox4{$#1\vcenter{}$}
	\rlap{\copy0}
	\dimen@=\dimexpr\ht2-\ht4-.2pt\relax
	\kern\dimen@
	{#2}%
	\kern\dimen@
	\copy0 
}
\makeatother

\allowdisplaybreaks

\makeatletter
\def\widebreve{\mathpalette\wide@breve}
\def\wide@breve#1#2{\sbox\z@{$#1#2$}%
     \mathop{\vbox{\m@th\ialign{##\crcr
\kern0.08em\brevefill#1{0.8\wd\z@}\crcr\noalign{\nointerlineskip}%
                    $\hss#1#2\hss$\crcr}}}\limits}
\def\brevefill#1#2{$\m@th\sbox\tw@{$#1($}%
  \hss\resizebox{#2}{\wd\tw@}{\rotatebox[origin=c]{90}{\upshape(}}\hss$}
\makeatletter

\usepackage{xr}
\usepackage{mathrsfs}

\makeatletter
\newcommand*{\addFileDependency}[1]{
  \typeout{(#1)}
  \@addtofilelist{#1}
  \IfFileExists{#1}{}{\typeout{No file #1.}}
}
\makeatother

\newcommand*{\myexternaldocument}[1]{
    \externaldocument{#1}
    \addFileDependency{#1.tex}
    \addFileDependency{#1.aux}
}

\myexternaldocument{noncompliancesupp15}

\usepackage{geometry}
\usepackage{setspace}
\begin{document}

\title{
	Complier General Causal Effect in Randomized Controlled Trials with One-Sided Noncompliance
	}

\author[a]{Yin Tang\thanks{e-mail: \texttt{yin.tang@uky.edu}}}
\author[b]{Yanyuan Ma\thanks{e-mail: \texttt{yzm63@psu.edu}}}
\author[c]{Jiwei Zhao\thanks{e-mail: \texttt{jiwei.zhao@wisc.edu}}}
\affil[a]{University of Kentucky}
\affil[b]{Pennsylvania State University}
\affil[c]{University of Wisconsin-Madison}

\date{\today}

\maketitle

\begin{abstract}
\noindent
A randomized controlled trial (RCT) is widely regarded as the gold
standard for assessing the causal effect of a treatment or
intervention, assuming perfect implementation.  
In practice, however, randomization can be compromised for various reasons, such as one-sided noncompliance.
In this paper, we first systematically study the likelihood-based identifiability in an RCT with one-sided noncompliance.
This foundational analysis naturally gives rise to the complier
general causal effect (CGCE) as the primary estimand.
We further develop two estimators for the CGCE: a simple estimator that requires no nonparametric procedures, and an efficient estimator that achieves the semiparametric efficiency bound. 
Our theoretical analysis shows that, 
achieving semiparametric efficiency requires only the nuisance estimators to converge in $L_2$-norm, with no restriction on their convergence rates. 
This rate-free property opens the door to employing many more modern machine learning methods while still guaranteeing efficiency.
Comprehensive simulation studies and a real data application are
conducted to illustrate the proposed methods and to compare them with
existing approaches. 
\end{abstract}

\newpage

\section{Introduction}\label{sec:intro}

A randomized controlled trial (RCT) is considered the gold standard for assessing the causal effect of a treatment or intervention, if perfectly implemented. 
In practice, however, randomization can be compromised due to complexities such as missing outcomes, dropout, or noncompliance \citep{follmann2000effect, mealli2004analyzing, dunn2005estimating, van2007estimation, hu2022assessing, zhang2023estimating}. 
Participants in RCTs, whether in biomedical or sociological contexts, often deviate from their assigned treatment and opt for a different one. 
In many settings, noncompliance is one-sided. 
For instance, in trials testing a new medical drug, individuals assigned to the control group typically cannot access the drug, whereas those assigned to treatment may choose not to take it. 
A similar pattern arises in training program evaluations: individuals assigned to the training may choose not to attend, while those assigned to the control group are generally unable or ineligible to participate.
In such scenarios, an intention-to-treat (ITT) analysis
\citep{frangakis1999addressing}, which evaluates outcomes based on
treatment assignment regardless of compliance, estimates the effect of
being assigned to treatment, rather than the actual causal effect of
receiving the treatment.

The challenge in estimating causal effects under noncompliance lies in the fact that the causal effect for the entire population is not identifiable from the observed data.
\cite{frangakis2002principal} introduced the principal stratification
framework, which partitions the study population into principal
strata, subpopulations defined by the joint potential compliance
behavior under alternative treatment assignments.  
The causal effect within each stratum, known as the principal causal
effect, is causally interpretable and can, under certain conditions,
be identified from the observed data.

In the literature, estimating causal effect with the issue of noncompliance has been studied
across disciplines, with researchers approaching it from
various perspectives.
For the average causal effect, the first set of results on the
identification and estimation of complier
average causal effect (CACE), also known as the local average treatment effect (LATE), was provided
by \cite{imbens1994identification}.  
Building on the instrumental variable (IV) framework introduced in \cite{angrist1996identification}, subsequent work has leveraged treatment assignment as an IV for the treatment received to obtain valid estimates of CACE. 
This line of research has since produced a rich literature under varying assumptions and settings, for example, \cite{abadie2003semiparametric}, \cite{tan2006regression}, \cite{frolich2007nonparametric}, \cite{wang2021estimation}, \cite{levis2024nonparametric}, \cite{baker2024multiple}, among others.
One can also refer to some textbooks, e.g., \cite{imbens2015causal},
for a comprehensive review of this topic. 
In contrast, the quantile causal effect \citep{doksum1974empirical, firpo2007efficient} has received considerably less attention in the presence of noncompliance.
\citet{wei2021estimation} investigated the complier
quantile causal effect (CQCE) for censored data with a binary IV.
More importantly, existing methods are limited to estimating a \emph{single} type of complier causal effect, either the CACE or the CQCE, rather than accommodating a more general causal estimand.

In this paper, we propose a novel estimand, complier
general causal effect (CGCE), in an RCT under one-sided noncompliance.
Its formal definition is presented in Section~\ref{sec:setting}.  
Note that CGCE is a general estimand, with some special cases including CACE and CQCE.
The motivation of proposing such a new estimand stems from our systematical study of the likelihood-based identifiability under this setting.
By constructing the nonparametric likelihood~\eqref{eq:model} from the data structure in Table~\ref{tab:onesided}, we clearly delineate which estimands are identifiable and which are not. 
To the best of our knowledge, this is the first systematic treatment of likelihood-based identifiability in this setting. 
This foundational analysis, with details presented in Section~\ref{sec:model}, naturally gives rise to the CGCE as the primary estimand.

We further develop two estimators for the CGCE: a simple estimator (Section~\ref{sec:simple}) that requires no nonparametric procedures and serves as a convenient tool for preliminary analysis, and an efficient estimator (Section~\ref{sec:efficient}) that achieves the semiparametric efficiency bound. 
A distinguishing feature of our approach to deriving the efficient influence function (EIF) is that, we  start from the influence function of the simple estimator and then project it onto the tangent space of the semiparametric model. 
This yields a transparent decomposition of the EIF in terms of interpretable nuisance components, each with a clear statistical meaning tied to the data structure in Table~\ref{tab:onesided}. 
Our approach stands in contrast to many other works that followed the optimal instrumental function approach, such as \cite{hong2010semiparametric}.

A key finding in our theoretical analysis is that achieving semiparametric efficiency requires only $L_2$-consistency of the nuisance estimates, with no restriction on their convergence rates. 
This rate-free property, enabled by sample splitting, distinguishes our work from many existing literature, whose efficient estimators rely on product-rate requirements or specific individual-rate requirements on sieve or kernel methods with some other smoothness conditions. 
This rate-free property opens the door to employing many more modern
machine learning (ML) methods while still guaranteeing efficiency.

To wrap up the introduction, below is the structure of the paper.
In Section~\ref{sec:setting}, we introduce the definition of one-sided noncompliance and the assumptions we impose.
In Section~\ref{sec:model}, we present the nonparametric likelihood, establish identifiability, and formally define the CGCE.
The \emph{simple} estimator and the \emph{efficient} estimator are studied in Sections~\ref{sec:simple} and \ref{sec:efficient}, respectively.
We conduct simulation studies in Section \ref{sec:simulation} and
analyze a social economic data set in Section \ref{sec:app}.
The paper is concluded with some discussions in Section~\ref{sec:disc}.
Detailed
derivations, regularity conditions, and all the technical proofs are placed in the Supplementary
Materials.

\section{Problem Setup}\label{sec:setting}

\subsection{One-sided noncompliance}

We first introduce some concepts in an RCT
setting, where $Z$ is the binary treatment each subject is
randomly assigned
($Z=1$ assigned to treatment and $Z=0$ control), and $T$
is the binary treatment each subject actually receives
($T=1$ treatment received and $T=0$ control received).
We consider the noncompliance issue in general; that is, $Z\neq T$.
To rigorously describe this issue, one formally recognizes the
variable $T$ as a potential outcome.
We postulate two potential outcomes $T_1$ and $T_0$, where $T_1$
($T_0$) is the treatment that the subject would have received if s/he
is assigned $Z=1$ ($Z=0$). That is, $T=Z T_1 + (1-Z)T_0$.
With one-sided noncompliance, we have $T_0=0$ thus $T=ZW$ by
simplifying the notation $T_1$ as $W$. 
In the literature, subjects with $W=1$ are called compliers and $W=0$ nevertakers.

The technical challenge is that the compliance status $W$ is not
always observed, as shown in {\color{gray}gray} color in
Table~\ref{tab:onesided}. 
When assigned to treatment with $Z=1$, we have $T=W$ so $W$ is
essentially observed; e.g., the first and second rows in
Table~\ref{tab:onesided}. 
However, when assigned to control with $Z=0$, we must have $T=0$ but
$W$ could be either 1 or 0; e.g., the third and fourth rows in
Table~\ref{tab:onesided}. 

\begin{table}[!tbp]
	\centering
	\caption{Data structure under one-sided noncompliance. Values in
		parentheses and with $\times$ are not observable. Variable $Z$ is
		the binary treatment assigned, and $T$ is the binary treatment
		received. Subjects with $W=1$ are called compliers and $W=0$
		nevertakers.} 
	\begin{tabular}{ccccccc}
		\hline
		$Z$ & ${\color{gray}W}$ & $T=Z {\color{gray}W}$  & $Y_{1}$ & $Y_{0}$ & $\X$\\
		\hline
		1 & 1 & 1 &  $\checkmark$ &$\times$ & $\checkmark$\\
		1 & 0 & 0 &  $\times$& $\checkmark$ & $\checkmark$\\
		0 & {\color{gray}(1)} & 0  &$\times$ & $\checkmark$ & $\checkmark$\\
		0 & {\color{gray}(0)} & 0  & $\times$& $\checkmark$ & $\checkmark$\\
		\hline
	\end{tabular}\label{tab:onesided}
\end{table}

\subsection{Assumptions and notation}

We make the following standard assumptions.
\begin{assumption}\label{as:sutva}
	The stable unit treatment value assumption (SUTVA), in that there are
	no causal effects of one subject's treatment assignment on another
	subject's outcome. 
\end{assumption}

\begin{assumption}\label{as:exclusion}
	Exclusion restriction.  We assume $Y_{Z,T}=Y_T$; i.e., the
	potential outcome is a function of the treatment received only and
	it does not depend on the treatment assigned.
\end{assumption}

\begin{assumption}\label{as:obs}
	Observed potential outcome assumption. We assume
	$Y=T Y_{1}+(1-T)Y_{0}$.
\end{assumption}

Assumptions~\ref{as:sutva}-\ref{as:obs} are all standard in causal inference.
Besides the potential outcome, we also assume the baseline covariate
$\X$ is available for every subject, and it follows the marginal
distribution $\fx(\x)$. 
We further assume
\begin{assumption}\label{as:randomization}
	$Z\indep
	(W, Y_{1},Y_{0})  \mid \X$.
	That is, the randomization procedure is performed based on the
	covariate only, and both the compliance status and potential
	outcomes are not related to the randomized treatment given
	covariate.
\end{assumption}
Assumption~\ref{as:randomization} is equivalent to the standard no unobserved
confounder assumption in causal inference. It is
reasonable because both
compliance status and potential outcomes are inherent
characteristics of an individual and its dependence on randomization
status is fully explained by the covariate already. 
Accordingly, we denote
\bse
p(\x) &\equiv& \pr(Z=1\mid w,y_1,y_0,\x)=\pr(Z=1\mid \x), \mbox{ and }\\
q(\x) &\equiv& \pr(W=1\mid z,\x) =\pr(W=1\mid\x).
\ese
Throughout the paper, we treat the function $p(\x)$ as known, as in an RCT.
In certain cases, $p(\x)$ may reduce to a known constant.

\begin{proposition}\label{pro:mar}
	Under one-sided noncompliance,
	Assumption~\ref{as:randomization} implies
	\be
	&& Z \indep (Y_{1}, Y_{0})\mid (W,\X), \mbox{ and }\label{eq:Zconind}\\
	&& T \indep (Y_{1},Y_{0})\mid (W,\X). \label{eq:mar}
	\ee
\end{proposition}
The proof of Proposition~\ref{pro:mar} can be found in Supplement S1.
Relation~\eqref{eq:mar} means, given the covariate and the compliance
status, the potential outcomes are independent of the treatment
received.
In other words, the potential
outcomes do not depend on the treatment  received given the
personal 
feature of  an individual, which includes the covariate and the compliance
status of that individual.  If $W$ had been observed, we can view
$T$ as the treatment assignment and $W$ as a component in the
covariate and proceed with the standard causal inference procedure
without considering noncompliance issue. However, $W$ is not always
observed, hence the problem becomes much harder because it can be
viewed as a
problem of a combination of missing covariate and causal inference.

\section{Likelihood, Identifiability, and Estimand}\label{sec:model}

Given the complexity discussed above, not all estimands under one-sided noncompliance are identifiable.
To understand the model identifiability, we first form the likelihood function of a generic observation, say $(\x,z,t,y)$, in each case corresponding to the four rows in
Table~\ref{tab:onesided}.

In the first row of Table~\ref{tab:onesided}, we denote the conditional pdf of $Y$ given $\X$, $Z=1$, $W=1$ as
\bse
f_1(y,1,\x)\equiv f_{Y\mid Z=1,W=1,\X}(y,1,\x)=
f_{Y_1\mid Z=1,W=1,\X}(y,1,\x)=f_{Y_1\mid W=1,\X}(y,1,\x),
\ese
where the last equality dues to the relation~\eqref{eq:Zconind}, $f_1$ stands for the pdf of $Y_1$, and the 1 in the argument
stands for $W=1$.
Hence the likelihood is $f_1(y,1,\x)q(\x)p(\x)f_\X(\x)$.

The other three rows in Table~\ref{tab:onesided} involve the potential outcome $Y_0$.
Similarly, we denote
\bse
f_0(y,w,\x)\equiv f_{Y_0\mid Z=1, W, \X}(y,w,\x)=f_{Y_0\mid W,\X}(y,w,\x)
\ese
as the conditional pdf of $Y_0$ given $\X$ and the compliance status $W$.
Thus, in the second row of Table~\ref{tab:onesided}, the likelihood is $f_0(y,0,\x)\{1-q(\x)\}p(\x)f_\X(\x)$.
For the third and fourth rows,
we can have either $W=1$ or $W=0$, then the likelihood is
\bse
&&f_{Y\mid Z,T,\X}(y\mid Z=0,T=0,\x)\pr(T=0\mid Z=0,\x)\pr(Z=0\mid
\x)f_\X(\x)\\
&=&f_0(y,0,\x)\{1-q(\x)\}\{1-p(\x)\}f_\X(\x)
+f_0(y,1,\x)q(\x)\{1-p(\x)\}f_\X(\x).
\ese

Therefore, the likelihood function of one generic observation $(\X,Z,T,Y)$, denoted as $f_{\X,Z,T,Y}(\x,z,t,y)$, is
\be\label{eq:model}
&&f_\X(\x)
\{f_1(y,1,\x)q(\x)p(\x)\}^{zt}
[f_0(y,0,\x)\{1-q(\x)\}p(\x)]^{z(1-t)}\\
&&\times[f_0(y,0,\x)\{1-q(\x)\}\{1-p(\x)\}
+f_0(y,1,\x)q(\x)\{1-p(\x)\}]^{(1-z)}.\n
\ee
This is a nonparametric likelihood with five components: $f_1(y,1,\x)$, $f_0(y,0,\x)$, $f_0(y,1,\x)$,  $q(\x)$ and $\fx(\x)$.
Fortunately, our next result shows that this nonparametric likelihood function is identifiable; i.e., any two different sets of these five components, $f_1(y,1,\x)$, $f_0(y,0,\x)$, $f_0(y,1,\x)$, $q(\x)$, $\fx(\x)$ and
$\wt f_1(y,1,\x)$, $\wt f_0(y,0,\x)$, $\wt f_0(y,1,\x)$, $\wt q(\x)$, $\wt f_{\X}(\x)$, will result in different likelihood functions.
Its proof can be found in Supplement S2.

\begin{lemma}[Identifiability]\label{lem:iden}
	The nonparametric likelihood~\eqref{eq:model}, $f_{\X,Z,T,Y}(\x,z,t,y)$, is identifiable.
\end{lemma}

This result is critical.
It indicates that any parameter of interest that is a functional of these
five components is identifiable, and is estimable with an appropriate device.
In this paper, we propose the estimand CGCE, complier general causal effect, introduced in Section~\ref{sec:intro},
defined as
$\btau\equiv\btau_1-\btau_0$, where $\btau_k$ solves
\be\label{eq:tau}
\0&=&  E\{\u(Y_k,\btau_k)\mid W=1\}\n\\
&=& E[E\{\u(Y_k,\btau_k)\mid W=1, \X\}\mid W=1]\n\\
&=& \frac{\int \u(y,\btau_k)f_k(y,1,\x)dy q(\x)f_\X(\x)d\x}{\int q(\x)f_\X(\x)d\x}
\ee
for $k=0,1$. 
It is clear that, by choosing $u(Y_k,\tau_k)$ as $Y_k-\tau_k$, the CGCE reduces to the CACE and by choosing $u(Y_k,\tau_k)$ as $I_{\{Y_k\leq \tau_k\}}-\alpha$, the CGCE becomes to the CQCE at the $\alpha$-percentile, $0<\alpha<1$.
Whenever quantile causal effect is in the context, we assume that the distribution functions of the potential outcomes are continuous and not flat at the $\alpha$-percentile, so that the corresponding quantiles are well defined and unique.
We skip those detailed assumptions; see \cite{firpo2007efficient}.


In addition, one can verify that the general causal effect among the treated,
i.e., if we had defined $\btau$ by conditional on $T=1$ instead
of $W=1$, 
is also identifiable, with a special case studied in \cite{frolich2013identification}.
However, neither general causal effect among nevertakers
(replacing $\mid W=1$ by $\mid W=0$ in the definition of $\btau$)
nor among the controls (replacing $\mid W=1$ by $\mid T=0$)
is identifiable since the involved component $f_1(y,0,\x)$ is not
available from the model~\eqref{eq:model}.
Certainly, the causal effect for the entire population is not identifiable.

Above, we study the likelihood-based identifiability in an RCT with one-sided noncompliance. 
To the best of our knowledge, this is the first systematic treatment
of likelihood-based identifiability in this setting.  
Our analysis above delineates which estimands are identifiable under
one-sided noncompliance and which are not, providing foundational
clarity for subsequent inference. 

In the following, we turn our attention to estimating CGCE $\btau$. 
We assume there are $n$ independent and identically distributed (iid)
observations $(\x_i,z_i,t_i,y_i)$, $i=1,\ldots,n$, for the random
variable $(\X,Z,T,Y)$. 

\section{Simple Estimation of CGCE}\label{sec:simple}

We start with a \emph{simple} estimator for $\btau$.
By \emph{simple}, we mean that we do not
need to engage any nonparametric estimation or ML tools.
We first introduce some notation on marginal probabilities:
$\rho_Z\equiv\pr(Z=1)$,
$\rho_W\equiv\pr(W=1)$,
$\rho_{wz}\equiv\pr(W=w,Z=z)$, $w=0,1$, $z=0,1$.
Because the compliance status $W$ is missing when $Z=0$, one might
think that it is hard to estimate $\rho_W$ at first sight. 
However,
our result below shows that $\rho_W$
can be straightforwardly estimated using the knowledge of $p(\x)$ and
$T$. 
\begin{proposition}\label{eq:rhoW}
	Under one-sided noncompliance and
	Assumptions \ref{as:sutva}-\ref{as:randomization}, $\rho_W
	=E\{T/p(\X)\}$.
\end{proposition}
The proof of Proposition~\ref{eq:rhoW} is contained in Supplement S3.
Thus, one can estimate $\rho_W$ by $\wh\rho_W \equiv n^{-1}\sumi T_i/p(\X_i)$.
For other marginal quantities, one can straightforwardly derive
$\wh\rho_Z\equiv n^{-1}\sumi Z_i$, $\wh\rho_{11}\equiv n^{-1}\sumi T_i$,
as well as
$\wh \rho_{01}\equiv n^{-1}\sumi (1-T_i)Z_i$ dues to the fact that
$\rho_{01}=\pr(W=0,Z=1)=\pr(T=0,Z=1)$. 
\begin{proposition}\label{pro:tau12}
	Under one-sided noncompliance and
	Assumptions \ref{as:sutva}-\ref{as:randomization}, we have
	\be
	\0 &=& E\{\u(Y_1,\btau_1)/p(\X)\mid
	T=1\}\frac{\rho_{11}}{\rho_W}, \label{eq:tau1} \\
	\0&=& E[\u(Y_0,\btau_0)/\{1-p(\X)\}\mid Z=0] \frac{1-\rho_Z}{\rho_W} - E\left\{\u(Y_0,\btau_0)/p(\X)\mid Z=1, T=0\right\}\frac{\rho_{01}}{\rho_W}.\label{eq:tau0}\n\\
	\ee
\end{proposition}
We defer its proof in Supplement S3.
We can estimate $\btau_1$ by  solving
$\0 = \sumi t_i\u(y_{1i},\btau_1)/p(\x_i)$
and $\btau_0$ by  solving $\0
= \sumi \u(y_{0i},\btau_0)(1-z_i)/\{1-p(\x_i)\} - \sumi
\u(y_{0i},\btau_0)(z_i-t_i)/p(\x_i)$ accordingly, where we used $z_it_i=t_i$.
Further, the simple estimator we propose for $\btau$ is
\be\label{eq:simple-est}
\wh\btau_s&\equiv& \wh\btau_{s1}-\wh\btau_{s0}\\
&=& {\rm argzero}_{\btau_1}
\sumi \frac{t_i \u(y_{1i},\btau_1)}{p(\x_i)} 
-{\rm argzero}_{\btau_0}
\sumi \u(y_{0i},\btau_0)\left\{\frac{1-z_i}{1-p(\x_i)} -  \frac{z_i-t_i}{p(\x_i)}\right\},\n
\ee
where we use the subindex $_s$ to denote simple.
The simple estimator $\wh\btau_s$ is root-$n$ consistent with its
influence function stated below in Theorem \ref{th:simple}, and its
proof is provided in Section S3. 

\begin{theorem}\label{th:simple}
	Under one-sided noncompliance,
	Assumptions \ref{as:sutva}-\ref{as:randomization} and 
	the regularity Conditions C1-C3 in
	Supplement S3.3, the
	simple estimator $\wh \btau_s$ in \eqref{eq:simple-est}
	satisfies
	$n^{1/2}(\wh\btau_s-\btau)=n^{-1/2}\sumi
	\bphi_s\{\X_i,Z_i,T_i,T_iY_{1i},(1-T_i)Y_{0i}\}+o_p(1)$, where
	\be\label{eq:eff-p2}
	\bphi_s\{\x,z,t,ty_1,(1-t)y_0\}\equiv- \A_1 \u(y_1,\btau_1)
	\frac{t}{p(\x)}+ \A_0
	\u(y_0,\btau_0)\left\{\frac{1-z}{1-p(\x)} -
	\frac{z-t}{p(\x)}\right\}\n
	\ee
	is the corresponding influence function.
	Here,
	\be\label{eq:a1a0}
	\A_1&=&\left[E
	\left\{W\frac{\partial\u(Y_1,\btau_1)}{\partial\btau\trans}\right\}\right]^{-1}
	=\left[E
	\left\{\frac{T}{p(\X)}\frac{\partial\u(Y_1,\btau_1)}{\partial\btau\trans}\right\}\right]^{-1},\nonumber\\
	\A_0&=&\left[E\left\{W
	\frac{\partial\u(Y_0,\btau_0)}{\partial\btau\trans}
	\right\}\right]^{-1}
	=\left(E\left[\left\{\frac{1-Z}{1-p(\X)}-\frac{Z-T}{p(\X)}\right\}
	\frac{\partial\u(Y_0,\btau_0)}{\partial\btau\trans}
	\right]\right)^{-1}.\quad
	\ee
	Thus, when $n\to\infty$,
	\bse
	n^{1/2}(\wh\btau_s-\btau)\to N(0,
	E[\bphi_s\{\X,Z,T,TY_1,(1-T)Y_0\}^{\otimes2}])
	\ese
	in distribution.
\end{theorem}
For estimating the asymptotic variance, one only need to construct
\bse
\wh \A_1
&=&\left[\frac{1}{n}\sumi
\left\{\frac{t_i}{p(\x_i)}\frac{\partial\u(y_{1i},\btau_1)}{\partial\btau\trans}\right\}\right]^{-1},\\
\wh \A_0&=&\left[\frac1n\sumi \left\{\frac{1-z_i}{1-p(\x_i)}-\frac{z_i-t_i}{p(\x_i)}\right\}
\frac{\partial\u(y_{0i},\btau_0)}{\partial\btau\trans}
\right]^{-1}.
\ese

Next, we would like to investigate more sophisticated estimation
strategies in the pursuit of efficiency, based on the simple estimator
$\wh\btau_s$. 
Despite its simplicity, the influence function of $\wh\btau_s$
motivates us a family of mean-zero estimating equations that
correspond to a family of robust estimators for $\btau$. 
Further, we use the projection technique to derive the efficient
influence function (EIF) for estimating $\btau$, where the influence
function of $\wh\btau_s$ serves as a preliminary basis for the derivation. 

\section{Efficient Estimation of CGCE}\label{sec:efficient}

\subsection{Influence functions}
Since $E\{ Z - p(\X) \mid \X\}=0$, it is straightforward to see that, for any
function $\varphi(\x)$, the following quantity has mean zero,
\be\label{eq:robust}
\bphi_r\{\X,Z,T,TY_1,(1-T)Y_0\} 
=\bvarphi (\X) \{ Z - p(\X) \} + \bphi_s\{\X,Z,T,TY_1,(1-T)Y_0\}.
\ee
Thus, for any pre-specified function $\bvarphi(\x)$, one can solve the
empirical version of the above mean zero estimating equation to
propose a corresponding estimator of $\btau$.

A more interesting question is, what is the \emph{optimal} choice of
$\varphi(\x)$ in the sense of estimation efficiency. 
By deriving the EIF for estimating $\btau$, we realize that the EIF
belongs to the family \eqref{eq:robust}.
Thus, the EIF is the best possible element in \eqref{eq:robust}.

To derive the EIF, 
we can engage the semiparametric tools \citep{bcrw, tsiatis}
to project the simple estimator's influence function $\bphi_s$ in
\eqref{eq:eff-p2} to the semiparametric tangent space. 
More specifically, we will derive the semiparametric tangent space
$\calT$, 
and then derive the EIF,
i.e., the projection of $\bphi_s$ onto the space $\calT$,
$\Pi(\bphi_s\mid \calT)$, in Proposition~\ref{pro:eff}. 
These derivations are technical and by no means trivial.
Readers of further interest can refer to Supplement S4.1
for the details.

\begin{proposition}\label{pro:eff}
	Under one-sided noncompliance and Assumptions
	\ref{as:sutva}-\ref{as:randomization}, the EIF for estimating $\btau$
	is 
	\be\label{eq:eff}
	\bphi\eff\{\X,Z,T,TY_1,(1-T)Y_0\}
	=\bphi_1 (\X) \{ Z - p(\X) \} + \bphi_s\{\X,Z,T,TY_1,(1-T)Y_0\},
	\ee
	where
	\be
	\bphi_1(\X)=
	\frac{\bmu_1(\X,\btau_1)q(\X)}{p(\X)}+
	\frac{\bmu_3(\X,\btau_0)}{1-p(\X)}+\frac{\bmu_2(\X,\btau_0)\{1-q(\X)\}}{p(\X)},\label{eq:eff-p1}
	\ee
	where
	\be
	\bmu_1(\X,\btau_1)=E\{\A_1\u(Y_1,\btau_1)|W=1,\X\}
	=E\{\A_1\u(Y_1,\btau_1)|Z=1,W=1,\X\}\label{eq:definemu1}
	\ee
	is the outcome corresponding to the first row in Table~\ref{tab:onesided},
	\be
	\bmu_2(\X,\btau_0)=  E\{\A_0\u(Y_0,\btau_0)\mid
	W=0,\X\}
	= E\{\A_0\u(Y_0,\btau_0)\mid Z=1,
	W=0,\X\},\label{eq:definemu2}
	\ee
	is the outcome corresponding to the second row in Table~\ref{tab:onesided}, and
	\be
	\bmu_3(\X,\btau_0)=E\{\A_0\u(Y_0,\btau_0)\mid\X\}
	=E\{\A_0\u(Y_0,\btau_0)\mid Z=0,\X\}\label{eq:definemu3}
	\ee
	is the outcome mean corresponding to the combination of the third and fourth rows in Table~\ref{tab:onesided}.
\end{proposition}
Clearly the EIF is indeed an element in \eqref{eq:robust} with the special choice of $\varphi(\x)$ shown in \eqref{eq:eff-p1}.
We note that the method we use here to derive the
EIF for CGCE is related to, but
distinct from, the method by
\cite{hong2010semiparametric}, which is based on the local instrumental
variable assumption. 
The nuisance components $\bmu_1, \bmu_2, \bmu_3$ and $q$ in our approach are
directly tied
to the data structure in Table~\ref{tab:onesided} and are designed to
handle much more generic estimands.

\subsection{Efficient estimator $\wh\btau$}\label{sec:eff}

Based on the EIF,
we  would like to construct the estimator $\wh\btau$ by solving
\bse
\sumi \bphi\eff\{\x_i,z_i,t_i,ty_{1i},(1-t)y_{0i}\}=\0.
\ese
In terms of implementation, one may opt to solve $\wh\btau_1$ and $\wh\btau_0$ separately, and then formulate $\wh\btau=\wh\btau_1 - \wh\btau_0$.
To this end, we show
in Section S4.2 of the supplement 
that the EIF for $\btau_1$ is
\bse
- \frac{\A_1 \u(Y_1,\btau_1) T}{p(\X)} + \frac{\bmu_1(\X,\btau_1) q(\X)}{p(\X)}\{Z-p(\X)\},
\ese
where $\bmu_1(\x,\btau_1)$ is defined by \eqref{eq:definemu1}.
This allows us to solve for $\wh\btau_1$ from
\be\label{eq:efftau1}
\0&=&\frac{1}{n}\sumi
\left[ -\frac{\wh \A_1 \u(y_{1i},\btau_1) t_i}{p(\x_i)} + \frac{
	\wh \bmu_1(\x_i,\btau_1)\wh q(\x_i)}{p(\x_i)}\{z_i-p(\x_i)\} \right],
\ee
where 
$\wh \bmu_1(\x_i,\btau_1) = \wh \A_1 \wh E \{\u(Y_{1i},\btau_1)|z_i=1,w_i=1,\x_i\}$.
Similarly, $\wh\btau_0$ can be obtained by solving
\be\label{eq:efftau}
\0
&=&\frac{1}{n}\sumi  \left( \wh \A_0
\u(y_{0i},\btau_0)\left\{\frac{1-z_i}{1-p(\x_i)} -
\frac{z_i-t_i}{p(\x_i)}\right\}\right.
\n\\
&&\left. +\left[
\frac{\wh\bmu_3(\x_i,\btau_0) }{1-p(\x_i)}
+\frac{\wh\bmu_2(\x_i,\btau_0)\{1-\wh q(\x_i)\}}{p(\x_i)}\right]\{ z_i - p(\x_i) \} \right),
\ee
where
\bse
\wh\bmu_2(\x_i,\btau_0) &=& \wh \A_0 \wh
E\{\u(Y_{0i}, \btau_0) \mid z_i=1,
w_i=0,\x_i\},\\
\wh\bmu_3(\x_i,\btau_0) &=& \wh
\A_0 \wh E\{\u(Y_{0i}, \btau_0)\mid z_i=0,\x_i\},\\
\wh q(\x_i) &=& \wh E(W_i\mid z_i=1,\x_i).
\ese
In practice, with any ML methods, one can estimate $\wh q(\x)$ by regressing 
$W$ on $\X$ based on the subgroup of the data $(W_i, \X_i, Z_i=1), i=1, \dots, n$. 
Similarly, one can estimate $\wh\bmu_k$'s by regressing $\u(Y_1,\btau_1)$ or $\u(Y_0,\btau_0)$ on $\X$ based on different subgroups of data: $\bmu_1(\x,\btau_1)$ with the subgroup $(Y_i, \X_i, T_i=1)$,
$\bmu_2(\x,\btau_0)$ with the subgroup  $(Y_i, \X_i, T_i=0, Z_i=1)$, and
$\bmu_3(\x,\btau_0)$ with the subgroup $(Y_i, \X_i, Z_i=0)$, for $i=1, \dots, n$.

Following the standard practice and to facilitate the theoretical
analysis, we
implement the estimator $\wh\btau$ via sample splitting.
Specifically,
we use the first  $n_0$ observations to estimate  $\wh\bmu_k$'s and
$\wh q$  and $\wh \A_1, \wh \A_0$, and use the remaining  $n_1$ observations to compute 
$\wh\btau = \wh\btau_1 - \wh\btau_0$. 
Here, $n=n_0+n_1$, and we  choose
$n_0=n_1=\lfloor n/2 \rfloor$ for convenience.
Denote  $\wh\bmu_{r1}, \wh\bmu_{r2}, \wh\bmu_{r3}, \wh q_r, \wh \A_{r1}, \wh \A_{r0}$ the corresponding estimates of
$\bmu_1, \bmu_2, \bmu_3, q, \A_1, \A_0$ based on the  $(r+1)$th part of
the data,  where $ 
r=0,1$. Note that in $\wh \A_{r1}, \wh \A_{r0}$, we can also plug in  initial estimators of $\btau_0,\btau_1$, 
such as the simple estimators $\wh\btau_{s0}$ and $\wh\btau_{s1}$. Then, for each $r$, we first solve $\wh\btau_{1-r,1}$ by solving \eqref{eq:efftau1}, 
with $\wh \A_1$, $\wh\bmu_1$ and $\wh q$ replaced by  $\wh \A_{r1}$, $\wh\bmu_{r1}$ and $\wh q_r$, respectively;
and we then  obtain the estimate of $\wh\btau_{1-r,0}$ by solving
\eqref{eq:efftau} with
$\wh \A_0$, $\wh\bmu_2$, $\wh\bmu_3$ and $\wh q$
replaced by $\wh \A_{r0}$, $\wh\bmu_{r2}$,
$\wh\bmu_{r3}$ and $\wh q_r$, respectively.
Let the estimate be $\wh\btau_{(r)} = \wh\btau_{1-r,1} - \wh\btau_{1-r,0}$.
Finally, we combine $\wh\btau_{(0)}$ and $\wh\btau_{(1)}$ to get $\wh\btau=(\wh\btau_{(0)}+\wh\btau_{(1)})/2$ as our final estimator.
We further denote
\be\label{eq:delta-def}
\bdelta_{r1}(\x,\btau_1) \equiv \wh \bmu_{r1}(\x,\btau_1) - \bmu_1(\x,\btau_1), &&
\bdelta_{r2}(\x,\btau_0) \equiv \wh \bmu_{r2}(\x,\btau_0) - \bmu_2(\x,\btau_0), \n\\
\bdelta_{r3}(\x,\btau_0) \equiv \wh \bmu_{r3}(\x,\btau_0) - \bmu_3(\x,\btau_0), &&
\delta_{rq}(\x) \equiv \wh q_r(\x) - q(\x).
\ee
Theorem \ref{th:eff} below shows that the estimator $\wh\btau$ defined above
indeed is the efficient estimator.
Its proof is contained in Supplement S4.3.

\begin{theorem}\label{th:eff}
	Under
	Assumptions \ref{as:sutva}-\ref{as:randomization} and 
	the regularity Conditions C1-C8  in
	Supplement S3.3. Assume
	the estimators $\wh \A_{rk}$, $\wh\bmu_{rk}(\cdot)$ and $\wh q_{r}(\cdot)$ satisfy
	\be\label{eq:pknown-tau-eff-cond}
	&&\wh \A_{r1} - \A_1 = o_p(1), \quad \wh \A_{r0} - \A_0 = o_p(1), \quad 
	E_\X \{ \delta_{rq}^2(\X) \} = o_p(1), \quad \n\\
	&& E_{\X} \left\| \bdelta_{r1}(\X,\btau_1) \right\|^2 = o_p(1), \quad 
	E_{\X} \left\| \bdelta_{rk}(\X,\btau_0) \right\|^2 = o_p(1), \, k=2,3,
	\ee
	where $\bdelta_{rk}$'s are defined in
	\eqref{eq:delta-def}.  Then, the estimator $\wh\btau$
	satisfies
	\bse
	n^{1/2}(\wh\btau-\btau)=n^{-1/2}\sumi
	\phi\eff\{\X_i,Z_i,T_i,T_iY_{1i},(1-T_i)Y_{0i}\}+o_p(1),
	\ese
	where  $\phi\eff(\cdot)$ is the EIF in \eqref{eq:eff}.
	Thus, when $n\to\infty$,
	\bse
	n^{1/2}(\wh\btau-\btau)\to N(0,
	E([\phi\eff\{\X,Z,T,TY_1,(1-T)Y_0\}]^2)
	\ese
	in distribution.
	Consequently, $\wh \btau$ is efficient.
\end{theorem}


\begin{remark}[Rate-free condition \eqref{eq:pknown-tau-eff-cond} in
	Theorem~\ref{th:eff}]
	A notable feature of Theorem~\ref{th:eff} is that it requires
	only $\bdelta_{rk}$ to converge to zero in second moment
	for $k=1,2,3,q$ and $r=0,1$, without imposing any specific
	convergence rate on the nuisance estimators. This rate-free
	condition is considerably weaker than the product-rate or
	individual-rate requirements commonly encountered in the
	semiparametric efficiency literature \citep{farrell2015robust, chernozhukov2018double,bonvini2024doubly}, where convergence rates
	tied to smoothness assumptions, sieve dimensions, or bandwidth
	parameters are typically needed.
	See also, for example, \cite{kennedy2016semi,ahrens2026introduction} for discussions about the rate requirements.
	
	From a practical standpoint, the rate-free condition
	dramatically broadens the class of methods available for
	estimating the nuisance functions $\A_k$, $k=0,1$,
	$\bmu_k$, $k=1,2,3$, and $q$. In high-dimensional settings,
	one may employ deep neural networks, classification and
	regression trees, random forests, and other modern ML methods whose theoretical properties may not be fully
	characterized beyond consistency, while still guaranteeing the
	semiparametric efficiency of $\wh\btau$. In lower-dimensional
	settings, more traditional nonparametric methods such as kernel
	regression or splines can also be used, and in such cases sample
	splitting may not even be necessary to achieve efficiency.
\end{remark}

In our simulation studies in Section~\ref{sec:simulation} and real data application in Section~\ref{sec:app}, we estimate the nuisance parameters $q$ and $\bmu_k$, $k=1,2,3$, by performing both kernel regressions and deep neural networks.  
For notation convenience, from now on, we define
$\sum_0$ by $\sum_{i=1}^{n_0}$ and
$\sum_1$ by $\sum_{i=n_0+1}^{n}$.
To be specific, in kernel estimators, for $r=0,1$, we use
\be\label{eq:kernel-nonpara-r}
\wh \bmu_{r1}(\x,\btau_1) &=& \wh \A_{r1}\frac{\sum_r Z_i W_i \u(Y_i,\btau_1) K_h (\x - \X_i)}{\sum_r Z_i W_i K_h (\x - \X_i)}, \n\\
\wh \bmu_{r2}(\x,\btau_0) &=& \wh \A_{r0}\frac{\sum_r Z_i (1-W_i) \u(Y_i,\btau_0) K_h (\x - \X_i)}{\sum_r Z_i (1-W_i) K_h (\x - \X_i)}, \n\\
\wh \bmu_{r3}(\x,\btau_0) &=& \wh \A_{r0}\frac{\sum_r (1-Z_i) \u(Y_i,\btau_0) K_h (\x - \X_i)}{\sum_r (1-Z_i) K_h (\x - \X_i)}, \n\\
\wh q_r(\x) &=& \frac{\sum_r Z_i W_i K_h (\x - \X_i)}{\sum_r Z_i K_h (\x - \X_i)},
\ee
and plug \eqref{eq:kernel-nonpara-r} back into \eqref{eq:efftau1},
\eqref{eq:efftau}
to compute the estimator. 
In the deep neural network based estimators, for each nonparametric component, we train a simple fully-connected neural network with ReLU activation function by minimizing the $L_2$-loss based on the corresponding group of the data. Specifically, denote the function class by NN, and we set
\be\label{eq:dnn-nonpara-r}
\wh \bmu_{r1}(\cdot,\btau_1) &=& {\arg\min}_{\f \in \mathrm{NN}} \sum_r Z_i W_i \| \u(Y_i, \btau_1) - \f (\X_i) \|^2, \n\\
\wh \bmu_{r2}(\cdot,\btau_0) &=& {\arg\min}_{\f \in \mathrm{NN}} \sum_r Z_i (1-W_i) \| \u(Y_i, \btau_0) - \f (\X_i) \|^2, \n\\
\wh \bmu_{r3}(\cdot,\btau_0) &=& {\arg\min}_{\f \in \mathrm{NN}} \sum_r (1-Z_i) \| \u(Y_i, \btau_0) - \f (\X_i) \|^2, \n\\
\wh q_r(\cdot) &=& {\arg\min}_{f \in \mathrm{NN}} \sum_r Z_i \{ W_i  - f(\X_i)\}^2,
\ee
and plug \eqref{eq:dnn-nonpara-r} back into \eqref{eq:efftau1},
\eqref{eq:efftau} to compute the estimator. The consistency of the deep neural networks are well investigated, for example, in \cite{schmidt2020nonparametric}. In our numerical studies, we conduct early-stopping to avoid overfitting. In fact, various types of neural networks as well as loss functions, can be applied to estimating the nonparametric components as long as they can produce a consistent estimator. The detailed implementation processes are provided in Section \ref{sec:simulation}.

\subsection{Discussions on $\wh\btau_s$ and $\wh\btau$}\label{sec:link}

In general, the influence function of the simple estimator $\wh\btau_s$, $\phi_s$, can be viewed as a \emph{special} $\phi\eff$ under misspecification,
where we  misspecify $\bmu_1(\X)$, $\bmu_2(\X)$ and
$\bmu_3(\X)$ to be $\0$.
This indicates that the efficient estimator is
robust in that we can misspecify many  terms in it, while under one
particular misspecification, we get the simple estimator. It also
indicates that
the simple estimator is
not efficient.
Below, Corollary~\ref{cor:ineff} verifies the inefficiency of simple estimator directly, with its proof contained in
Supplement S5.

\begin{corollary}\label{cor:ineff}
	Under 
	Assumptions \ref{as:sutva}-\ref{as:randomization} and 
	the regularity Conditions C1-C2 in Supplement
	S3.3, 
	the simple estimator
	$\wh \btau_s$ in \eqref{eq:simple-est} is not efficient, i.e.,
	$ \lim_{n\to\infty} n \var(\wh\btau_s)= E (\bphi_s^{\otimes 2})>
	E (\bphi\eff^{\otimes 2})$, where $\A > \B$ means that $\A-\B$ is a
	positive definite matrix. 
\end{corollary}

Although the simple estimator is not fully efficient, it has the
advantage of being \emph{simple} in that it does not require any
nonparametric procedures,  hence is a convenient tool to obtain
preliminary analysis.

Finally we consider a special case in which the covariate $\X$ is absent. 
This scenario commonly appears in classic textbooks when introducing
the concept of one-sided noncompliance; e.g., \cite{imbens2015causal}. 
Under such a scenario, one can verify that the efficient estimator of CACE takes the 
explicit form: 
\be\label{eq:simple-degenerate}  
\left\{ \frac{\sumi y_iz_i}{\sumi z_i} - \frac{\sumi y_i(1-z_i)}{\sumi (1-z_i)} \right\} / \left(\frac{\sumi t_i}{\sumi z_i} \right), 
\ee 
which coincides with the estimator originally appeared in the literature; see, for example, Chapter 23 in
\cite{imbens2015causal}.
This analysis reveals that, in the absence of $\X$, the commonly used estimator \eqref{eq:simple-degenerate} is
already efficient.

\section{Simulation Studies}\label{sec:simulation}

In this section,  we perform simulation studies to evaluate the
performance of the  two
estimators  for the complier average causal effect, 
where we consider $u(y,\tau) = y-\tau$. We consider two scenarios.
In the first scenario, we consider the dimensions of $\X$ to be
$d=1,4,9$, and  the data sets are generated as below.
For each $d$, we first form $\X$ by independently  generating each
component of $\X$ from
Uniform$(1,5-\sqrt{d})$. Let $X_0 = \sum_{j=1}^d X_j$ for later convenience.
Given $\X$, we generate $Z \sim \mathrm{Bernoulli}\{p(\X)\}$ and $W \sim \mathrm{Bernoulli}\{q(\X)\}$ independently, where
\bse
p(\x) = \frac{1}{4} \sin ( \pi x_0) + \frac{1}{2}, \quad q(\x) = \frac{1}{4} \cos ( 2\pi x_0) + \frac{1}{2}.
\ese
We further set $T=ZW$.
We also generate $\epsilon \sim N(0,1)$ independently of $\X$, $Z$ and $W$, and we set
\bse
Y_1 = 2 + 4 X_0 + \epsilon, \quad Y_0 = 1 + 2 X_0 + \epsilon.
\ese
The observed response is $Y = T Y_1 + (1-T) Y_0$ by definition. Note
that $W$ is not observed,  so the observations are $(\X_i, Z_i,
T_i, Y_i)$, for $i=1, \dots, n$. We set $n=10,000$, and
repeat the simulation 1,000 times.

Under the above setting,  we can obtain that
$\mu_1(\x) = 2 + 4 x_0$,
$\mu_2(\x) = 1 + 2 x_0$,
$\mu_3(\x) = 1 + 2 x_0$, and
$\btau =1 + 2 E (X_0 \mid W=1)$.
Note that
\bse
E (X_0 \mid W=1) = \frac{\int x q(x) f_{X_0}(x) dx}{\int q(x) f_{X_0}(x) dx}
\ese
where $f_{X_0}(x) = \frac{1}{4-\sqrt{d}}
f_U\left(\frac{x-d}{4-\sqrt{d}}\right)$, and $f_U(u) =
\frac{1}{(d-1)!} \sum_{k=0}^{\lfloor u \rfloor} (-1)^k {d \choose k}
(u-k)^{d-1}$ is the density of the Irwin–Hall distribution. By numerical integration, we
get the true values of $\btau$ to be  $6, 17, 28$ when
$d=1,4,9$, respectively.

We  implemented kernel regression for $\mu_1, \mu_2, \mu_3, q$ in
the efficient
estimator  for all dimensions, and also  implemented
deep neural network  for dimensions $d=4$ and $d=9$.
We also implemented the oracle estimator by adopting the true
functions $\mu_1, \mu_2, \mu_3, q$ in the efficient estimator implementation.
When implementing the kernel estimators, we use the product of
one-dimensional kernels  for all $d$, where we choose the
one-dimensional kernel function to be
$K(x) = \phi(x)$ when $d=1$,
$K(x) = (15-10x^2+x^4)\phi(x)/8$ when $d=4$, and
$K(x) = (945-1260x^2+378x^4-36x^6+x^8)\phi(x)/384$ when
$d=9$,
where $\phi$ is the pdf of standard normal distribution.
The bandwidth $\h$ is set as $h_j = 1.5
\sqrt{d} m^{-1/(2d+1)} \wh\sigma_j$, where  $m$ is the sample
size engaged for the particular kernel based estimation, and
$\wh\sigma_j$ is the
estimated standard deviation of $X_j$, for $j=1,\ldots, d$.
When implementing the deep neural network estimators, we construct a
4-layer fully-connected neural network with 512 neurons in each
layer. We use the mean squared error as the loss function,
and use Adams to perform the optimization,  with learning rate 0.01.

\begin{figure}[htbp]
	\centering
	\includegraphics[width=\textwidth]{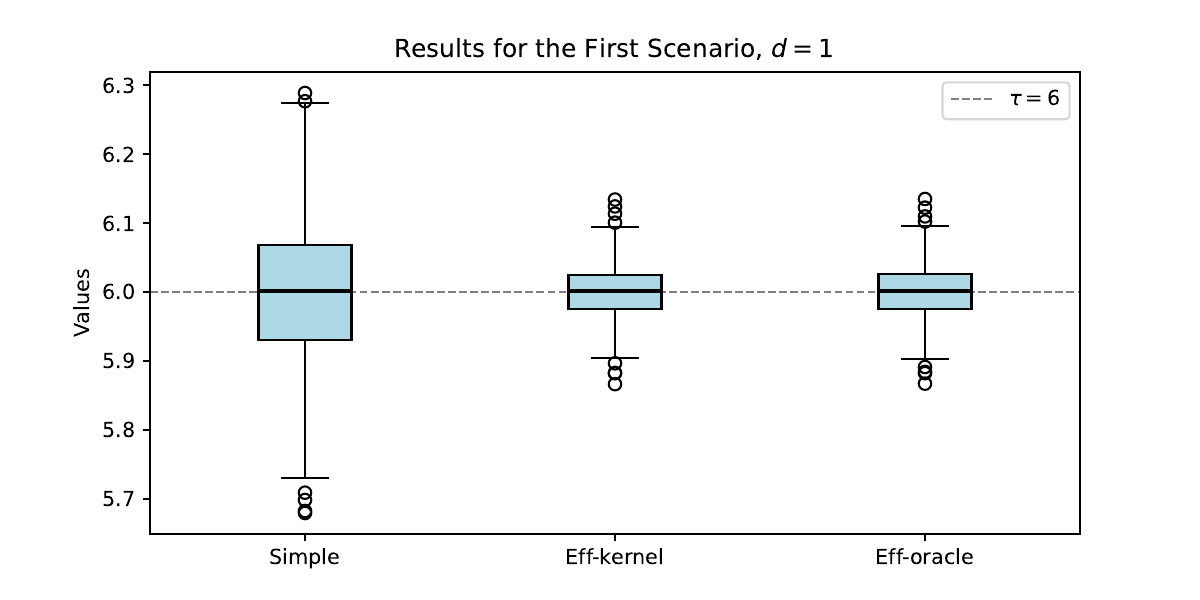}
	\caption{Boxplot of 1,000 estimates of $\btau$,
		first scenario, $d=1$.
		The horizontal line is the true CACE $\btau = 6$.
		Simple and efficient estimators are implemented,
		with nonparametric components estimated using kernel,
		as well as the oracle estimator.
	}
	\label{fig:res-1dim-s1}
\end{figure}

\begin{figure}[htbp]
	\centering
	\includegraphics[width=\textwidth]{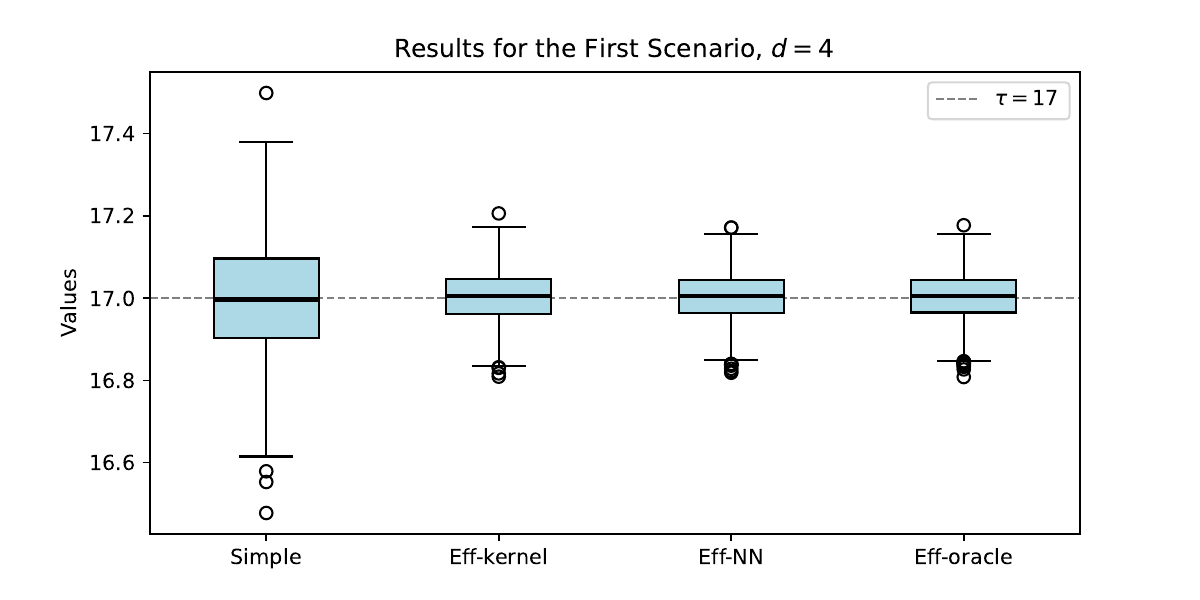}
	\caption{Boxplot of 1,000 estimates of $\btau$,
		first scenario, $d=4$.
		The horizontal line is the true CACE $\btau = 17$.
		Simple and efficient estimators are implemented,
		with nonparametric components estimated using kernel and neural network(NN),
		as well as the oracle estimator.}
	\label{fig:res-4dim-s1}
\end{figure}

\begin{figure}[htbp]
	\centering
	\includegraphics[width=\textwidth]{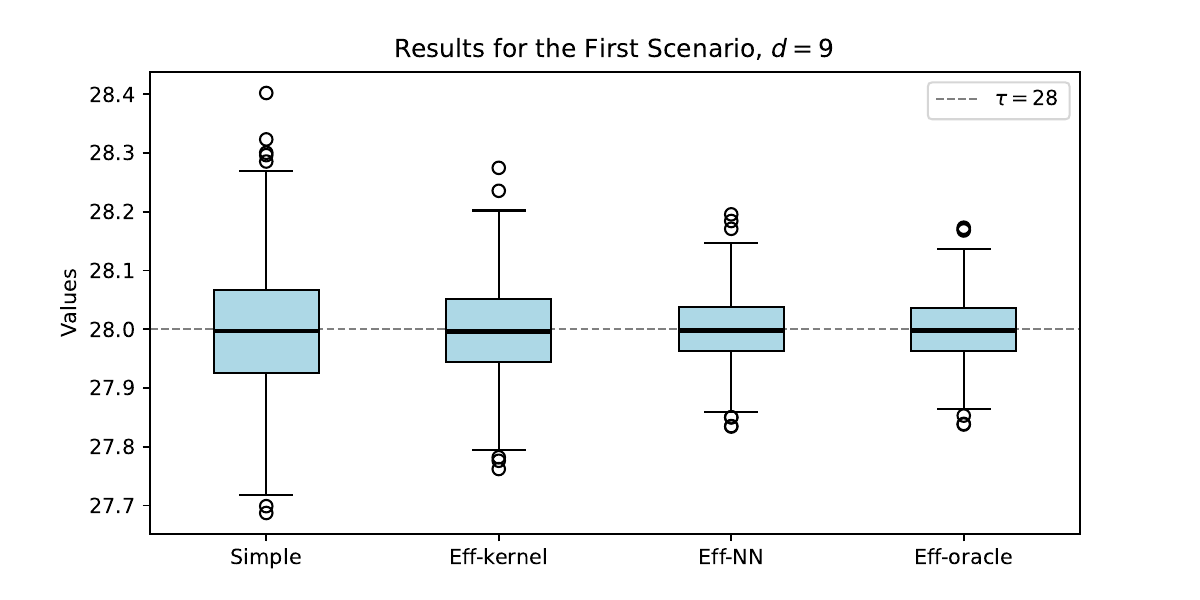}
	\caption{Boxplot of 1,000 estimates of $\btau$,
		first scenario, $d=9$.
		The horizontal line is the true CACE $\btau = 28$.
		All column indices have the same meaning as in Figure \ref{fig:res-4dim-s1}.}
	\label{fig:res-9dim-s1}
\end{figure}

To avoid overfitting, we adopt the early stopping criterion by
randomly splitting 20\% of the data as the validation set and using
the remaining 80\% as the training set. The stopping criterion is
either when the validation loss does not improve by a small $\delta$
within 10 steps, or the number of iterations reaches an upper bound we
set. Here, we set the $\delta$ to be $10^{-6}$ when estimating $q$,
and $10^{-4}$ when estimating $\mu_1,\mu_2,\mu_3$ due to the
difference of their ranges, and we set the maximum number of
iterations as 800. To avoid the effect of random initialization, we
also force the number of iterations to be at least 50. The results are
presented  in Figures \ref{fig:res-1dim-s1} to \ref{fig:res-9dim-s1}
and Tables \ref{tab:res-1dim-s1} to \ref{tab:res-9dim-s1}.

\begin{table}[htbp]
	\centering
	\begin{tabular}{lcccccc}
		\hline
		Method &    Mean &     Bias &      SD &    RMSE &  $\wh{\mathrm{SD}}$ & 95\% cvg \\
		\hline
		Simple & 5.9997 & -0.0003 & 0.103 & 0.103 &  0.105 &    0.958 \\
		Eff-kernel & 6.0004 &  0.0004 & 0.038 & 0.038 &  0.040 &    0.956 \\
		Eff-oracle & 6.0004 &  0.0004 & 0.038 & 0.038 &  0.040 &    0.957 \\
		\hline
	\end{tabular}
	\caption{Results based on 1,000 estimates of $\btau$,
		first scenario, $d=1$. True $\btau = 6$.
		Simple and efficient(Eff) estimators are implemented,
		with nonparametric components estimated using kernel,
		as well as the oracle estimator.
		$\wh {\rm SD}$ is computed based on
		the asymptotic result of the corresponding estimator, and
		95\% cvg is the empirical coverage of the 95\% asymptotic confidence intervals.}
	\label{tab:res-1dim-s1}
\end{table}

\begin{table}[htbp]
	\centering
	\begin{tabular}{lcccccc}
		\hline
		Method &    Mean &     Bias &      SD &    RMSE &  $\wh{\mathrm{SD}}$ & 95\% cvg \\
		\hline
		Simple & 16.9978 & -0.0022 & 0.143 & 0.143 &  0.137 &    0.935 \\
		Eff-kernel & 17.0026 &  0.0026 & 0.064 & 0.064 &  0.062 &    0.939 \\
		Eff-NN & 17.0036 &  0.0036 & 0.062 & 0.062 &  0.061 &    0.939 \\
		Eff-oracle & 17.0034 &  0.0034 & 0.060 & 0.060 &  0.059 &    0.946 \\
		\hline
	\end{tabular}
	\caption{Results based on 1,000 estimates of $\btau$,
		first scenario, $d=4$. True $\btau = 17$.
		Simple and efficient(Eff) estimators are implemented,
		with nonparametric components estimated using kernel and neural network(NN),
		as well as the oracle estimator.
		All column indices have the same meaning as in Table \ref{tab:res-1dim-s1}.}
	\label{tab:res-4dim-s1}
\end{table}

\begin{table}[htbp]
	\centering
	\begin{tabular}{lcccccc}
		\hline
		Method &    Mean &     Bias &      SD &    RMSE &  $\wh{\mathrm{SD}}$ & 95\% cvg \\
		\hline
		Simple & 27.9974 & -0.0026 & 0.108 & 0.108 &  0.106 &    0.941 \\
		Eff-kernel & 27.9981 & -0.0019 & 0.076 & 0.076 &  0.073 &    0.940 \\
		Eff-NN & 27.9995 & -0.0005 & 0.054 & 0.054 &  0.054 &    0.959 \\
		Eff-oracle & 27.9991 & -0.0009 & 0.052 & 0.052 &  0.052 &    0.954 \\
		\hline
	\end{tabular}
	\caption{Results based on 1,000 estimates of $\btau$,
		first scenario, $d=9$. True $\btau = 28$.
		All row and column indices have the same meaning as in Table \ref{tab:res-4dim-s1}.}
	\label{tab:res-9dim-s1}
\end{table}

Based on these results, in terms of estimation performance,
all estimators have very small bias,
suggesting the consistency. On the other hand,
the simple estimator has much larger variability than all other
estimators in all cases, reflecting our theory that the simple
estimator is not efficient. 
The efficient estimator has very small variability
regardless it is combined with kernel method or neural network method
for $d=1, 4$. When $d=9$, the advantage of the neural network method
starts to show, in that Eff-NN has smaller variability than Eff-kernel.
In terms of the inference performance, both simple
and efficient estimators perform very well, in that the estimated
standard deviation is close to the sample version, and the constructed
95\% confidence intervals indeed covers the truth about 95\% of the
times. 
It is worth noting that in all the settings, the performance
of the efficient estimator in combination with neural network
always performs closely to the oracle estimator,
which shows its superiority.

\begin{figure}[htbp]
	\centering
	\includegraphics[width=\textwidth]{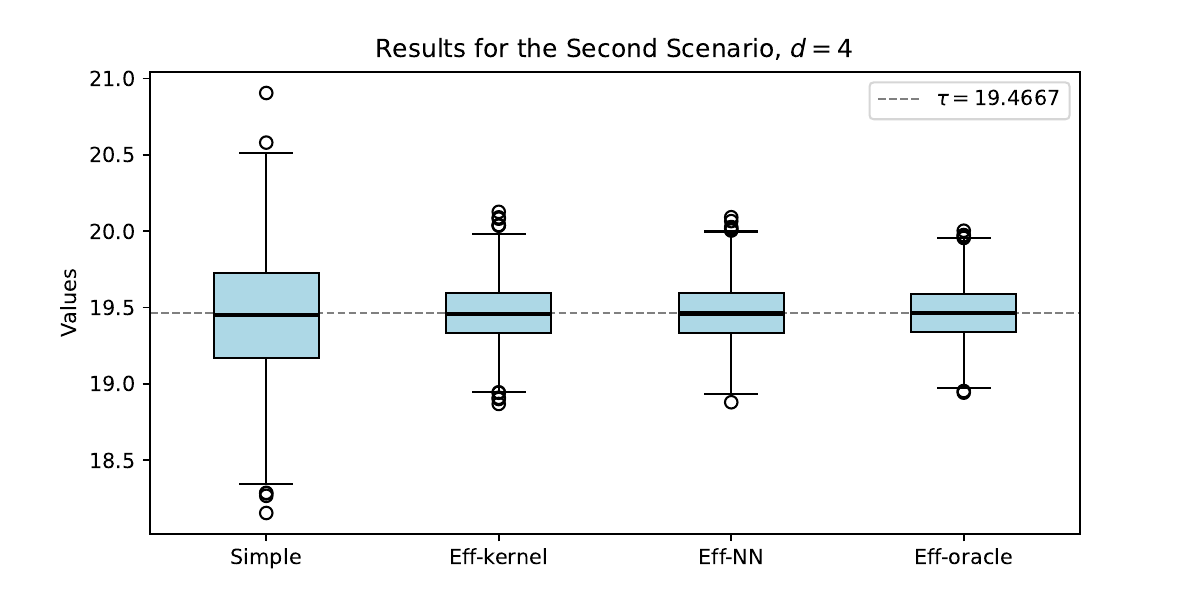}
	\caption{Boxplot of 1,000 estimates of $\btau$,
		second scenario, $d=4$.
		The horizontal line is the true CACE $\btau = 19.4667$.
		All column indices have the same meaning as in Figure \ref{fig:res-4dim-s1}.}
	\label{fig:res-4dim-s2}
\end{figure}

\begin{figure}[htbp]
	\centering
	\includegraphics[width=\textwidth]{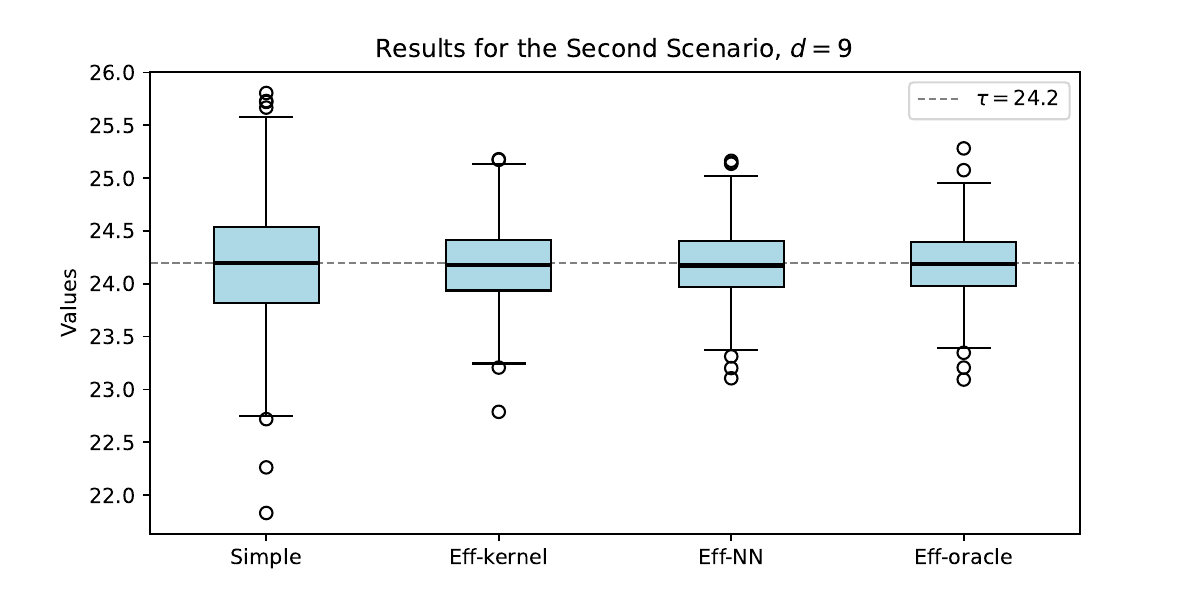}
	\caption{Boxplot of 1,000 estimates of $\btau$,
		second scenario, $d=9$.
		The horizontal line is the true CACE $\btau = 24.2$.
		All column indices have the same meaning as in Figure \ref{fig:res-4dim-s1}.}
	\label{fig:res-9dim-s2}
\end{figure}

In the second scenario, we only considered dimensions $d=4$ and $d=9$.
All the data generation procedures are identical to the first
scenario, except that
we generated $Y_1$ and $Y_0$ from
\bse
Y_1 = 2 + 2W+ (4 +2W) X_0 + 0.1 X_0^2 +  \epsilon, \quad Y_0 = 1 + W + (2 + W) X_0 + 0.2 X_0^2 + \epsilon.
\ese
This leads to  $\mu_1(\x)  = 4 + 6 x_0 + 0.1 x_0^2$,
$\mu_2(\x)  = 1 + 2 x_0 + 0.2 x_0^2$, $\mu_3(\x) = 1 + q(\x) + \{ 2 + q(\x) \} x_0 + 0.2 x_0^2$, and
$\btau = 2 + 3 E (X_0 \mid W=1) - 0.1 E (X_0^2 \mid W=1) $.
By numerical integral, the true $\btau$ is  $19.4667$ for $d=4$,
and
$24.2$ for $d=9$.
The results presented in Figures \ref{fig:res-4dim-s2},
\ref{fig:res-9dim-s2} and
Tables \ref{tab:res-4dim-s2}, \ref{tab:res-9dim-s2}
lead to the same conclusions as in the first scenario, hence we do not
repeat.

\begin{table}[htbp]
	\centering
	\begin{tabular}{lcccccc}
		\hline
		Method &    Mean &     Bias &      SD &    RMSE &  $\wh{\mathrm{SD}}$ & 95\% cvg \\
		\hline
		Simple & 19.4477 & -0.0190 & 0.413 & 0.414 &  0.415 &    0.947 \\
		Eff-kernel & 19.4636 & -0.0031 & 0.206 & 0.206 &  0.209 &    0.953 \\
		Eff-NN & 19.4649 & -0.0017 & 0.201 & 0.201 &  0.207 &    0.947 \\
		Eff-oracle & 19.4654 & -0.0012 & 0.189 & 0.189 &  0.192 &    0.951 \\
		\hline
	\end{tabular}
	\caption{Results based on 1,000 estimates of $\btau$,
		second scenario, $d=4$. True $\btau =19.4667$.
		All row and column indices have the same meaning as in Table \ref{tab:res-4dim-s1}.}
	\label{tab:res-4dim-s2}
\end{table}

\begin{table}[htbp]
	\centering
	\begin{tabular}{lcccccc}
		\hline
		Method &    Mean &     Bias &      SD &    RMSE &  $\wh{\mathrm{SD}}$ & 95\% cvg \\
		\hline
		Simple & 24.1833 & -0.0167 & 0.540 & 0.540 &  0.541 &    0.957 \\
		Eff-kernel & 24.1817 & -0.0183 & 0.356 & 0.356 &  0.357 &    0.955 \\
		Eff-NN & 24.1857 & -0.0143 & 0.317 & 0.317 &  0.319 &    0.952 \\
		Eff-oracle & 24.1860 & -0.0140 & 0.301 & 0.301 &  0.295 &    0.950 \\
		\hline
	\end{tabular}
	\caption{Results based on 1,000 estimates of $\btau$,
		second scenario, $d=9$. True $\btau = 24.2$.
		All row and column indices have the same meaning as in Table \ref{tab:res-4dim-s1}.}
	\label{tab:res-9dim-s2}
\end{table}

\section{Real Data Application}\label{sec:app}

We apply our methodology to analyze a microcredit data set from an
experiment conducted in Morocco.
The study aims to analyze the causal effect of microcredit on the
output from self-employment activities.
As described in \cite{sawada2019noncompliance}, Al Amana, a local
microfinance institute, opened new branches in some villages at the
beginning of the experiment, and the authors of
\cite{crepon2015estimating} conducted a
baseline survey on the households in 162 villages.
Based on the baseline survey,  they divided the villages into 81
pairs, each pair with
similar characteristics, and randomly  assigned one for treatment
and the other for control.  Thus, on the level
of the households, each household has 1/2 probability of receiving
treatment regardless of the household situation.
In the treatment villages,
the agents of Al Amana promoted participation in
microcredit, but the control villages did not have access to microcredit.
In the treatment villages, people could still choose either to  apply
microcredit or not.  This corresponds to the one-sided
noncompliance scheme. The study is conducted for 12 months,
and the response is considered as the total output from self-employment
activities  of a household during the time.

In the dataset, the covariates for each household are collected
at the baseline survey.
Similar to \cite{sawada2019noncompliance}, we  include $d=9$ covariates in
$\X$. These covariates include 3 continuous variables, the number of
household members, the number of adults (members 16 years old or
older), and the household head's age,  as well as 6 categorical variables,
the indicator variables for animal husbandry self-employment activity,
non-agricultural self-employment activity, outstanding loans borrowed
from any source, spouse or head respondence to self-employment
section, other member respondence to self-employment section, and the
missingness of the number of household members at baseline.

The treatment assignment mechanism leads to $Z=1$ with half
probability, i.e.  $p(\X) = 0.5$ for
any $\X$.
Let $W$ be whether or not a household follows the promoted microcredit
policy and
$T$ be whether an individual  received
microcredit. By the design, we have $T=0$ for all households in
control villages, while $T$  is either 0 or 1 for households in the
treatment villages.  Note that $T$ is available in the data set
while $W$ is not. The total output from self-employment
activities  of a household forms the response variable.
Same as \cite{sawada2019noncompliance}, we use a subsample of units
with high borrowing probabilities and endline observations, which
contains $n=4,934$ observations in total.

Following \cite{sawada2019noncompliance}, we apply the inverse hyperbolic
sine transformation $\log (y+\sqrt{y^2+1)}$ on the original total
output, and use the transformed output as our response $Y$.

Further, slightly different from \cite{sawada2019noncompliance},
we combine  the variable denoting ``the spouse or head respondence to
self-employment section'' and its missingness indicator variable
into a single variable with three values $\{-1,0,1\}$, where $-1$ means
missing, $0$ means no, and $1$ means yes.
We also standardize
the three continuous covariates.

To evaluate the performance of the various methods in this application,
we conduct a simulation study by drawing  $N=1,000$
bootstrap samples from the dataset, each  containing $n=4,934$
households, and perform the same analysis on each bootstrap
sample. We conduct the same analysis as in  the simulation studies with
$d=9$. We  implemented the  three
methods -- Simple, Eff-kernel and Eff-NN -- on the
bootstrap
datasets.  In estimating $q(\x)$ via DNN, we
used the cross-entropy loss.
For
each method, we use the estimate from the original dataset as the true
value, and compute the empirical coverage of the 95\% confidence
intervals.  Because there are some extreme values in the estimates of
kernel-based methods, we report the median of the estimates and the
median of the estimated standard deviations for the 1,000 bootstrap
samples,  as well as the sample standard deviation based on median
absolute deviation (MAD).  Here,
the MAD of $\wh
\btau_1,\dots, \wh \btau_N$ is defined as $\mathrm{median}\{ | \wh
\btau_i - \mathrm{median}(\wh \btau_j) | \}$, and the estimator is $1.4826
\mathrm{MAD}$
\citep{leys2013detecting}.
The results are summarized in Table \ref{tab:res-application}.

\begin{table}[htbp]
	\centering
	\begin{tabular}{lcccccc}
		\hline
		Method & Truth & Median & Bias &  SD & $\wh{\mathrm{SD}}$ & 95\% cvg \\
		\hline
		Simple & 1.425 &  1.468 &    0.043 & 0.670 &   0.749 &    0.969 \\
		Eff-kernel & 1.604 &  1.379 &   -0.226 & 0.923 &   0.791 &    0.926 \\
		Eff-NN & 1.113 &  1.239 &    0.126 & 0.658 &   0.691 &    0.952 \\
		\hline
	\end{tabular}
	\caption{Results based on 1,000 bootstrap samples in the real data analysis. 
		Bias is based on the median of estimators.
		SD is the MAD based estimator of the standard deviation based on the 1,000 estimates,
		$\wh {\rm SD}$ is the median of
		the asymptotic standard deviations of the corresponding estimator, and
		95\% cvg is the empirical coverage of the 95\% asymptotic confidence intervals.
	}
	\label{tab:res-application}
\end{table}

According to Table \ref{tab:res-application}, we see that SD and
$\wh{\mathrm{SD}}$ match well for
the efficient estimator combined with neural networks
(Eff-NN), and its empirical coverage of the 95\% confidence
interval is also very close to 0.95,
indicating its good finite sample inference result.
The SD and $\wh{\mathrm{SD}}$ are also reasonably close for the simple
estimator (Simple), 
and its empirical coverage of the 95\% confidence interval is slightly higher than 0.95.
However,  the efficient 
estimator in combination with kernel method underestimates the standard deviation.
This is  
because kernel-based methods do not  work well when the
dimension is high and the sample size is not  sufficiently large.

Based on the performance in
Table \ref{tab:res-application}, we will only perform inference
using the efficient method combined with neural networks (Eff-NN).
The efficient estimator yields $\wh \btau = 1.113$ with the estimated
standard deviation $\wh \sigma = 0.604$, leading to the asymptotic
95\% confidence interval $[-0.071,2.296]$.
On the other hand, we may also consider the simple method though it is
slightly conservative.
The simple estimator yields $\wh \btau = 1.425$ with the estimated
standard deviation $\wh \sigma = 0.742$.
The asymptotic 95\% confidence interval for $\btau$ is
$[-0.029,2.878]$.
Both confidence
intervals contain $0$, indicating that there is no significant
evidence to claim  $\btau$, the average treatment effect under
compliance,  is different from zero. 
These results are
different from the those in \cite{sawada2019noncompliance}.
We conjecture that this is because we do not make any parametric model
assumptions throughout the analysis, while
\cite{sawada2019noncompliance} adopts a linear model with the
treatment assignment and the treatment received as two dummy
variables.

\section{Discussion}\label{sec:disc}

In this paper, we propose a new estimand, CGCE, complier general causal effect, in an RCT under one-sided noncompliance.
We first systematically study the likelihood-based identifiability under this setting.
By constructing the nonparametric likelihood~\eqref{eq:model} from the data structure in Table~\ref{tab:onesided}, we clearly delineate which estimands are identifiable and which are not. 
To the best of our knowledge, this is the first systematic treatment of likelihood-based identifiability in this setting. 
This foundational analysis naturally gives rise to the CGCE as the primary estimand, a unified framework that encompasses both the CACE and CQCE as special cases.

We develop two estimators for the CGCE: a simple estimator that requires no nonparametric procedures and serves as a convenient tool for preliminary analysis, and an efficient estimator that achieves the semiparametric efficiency bound. 
A distinguishing feature of our approach to deriving the EIF is that, we  start from the influence function of the simple estimator and then project it onto the tangent space of the semiparametric model. 
This yields a transparent decomposition of the EIF in terms of interpretable nuisance components, each with a clear statistical meaning tied to the data structure in Table~\ref{tab:onesided}. 
Our approach stands in contrast to many other work that followed the optimal instrumental function approach, such as \cite{hong2010semiparametric}.

Perhaps our most striking theoretical finding is the rate-free condition identified in Theorem~\ref{th:eff}: achieving semiparametric efficiency requires only that the nuisance estimators converge to their true values in $L_2$-norm, with no restriction on their convergence rates. 
This is considerably weaker than the product-rate or individual-rate requirements commonly seen in the semiparametric efficiency literature. 
Enabled by sample splitting, this rate-free property opens the door to employing many more modern ML methods while still guaranteeing efficiency.

Finally, we remark on several promising directions for future work. 
The extension of our framework to two-sided noncompliance, where both always-takers and nevertakers are present, should be feasible; the main challenge lies in the more complex likelihood structure and the additional unidentifiable components that arise. 
The extension to observational data settings, where the propensity score is unknown and has to be estimated, is also a natural next step and would further broaden the applicability of the CGCE framework. 
We leave these extensions for future investigation.

\section*{Supplement}

The supplement includes all of the
derivations, regularity conditions, and all the proofs.

\section*{Acknowledgment}

The research is supported in part by NSF (DMS 1953526, 2122074, 2310942), NIH (R01DC021431) and the American Family Funding Initiative of UW-Madison.

\section*{Conflict of Interest}

The authors report there are no competing interests to declare.

\bibliographystyle{asa}
\bibliography{reference}
\end{document}